\documentclass[english,12pt,preprint]{aastex}
\usepackage[T1]{fontenc}
\usepackage[latin1]{inputenc}
\usepackage{babel}
\usepackage{graphics}

\makeatletter

\providecommand{\LyX}{L\kern-.1667em\lower.25em\hbox{Y}\kern-.125emX\@}
\let\SF@@footnote\footnote
\def\footnote{\ifx\protect\@typeset@protect
    \expandafter\SF@@footnote
  \else
    \expandafter\SF@gobble@opt
  \fi
}
\expandafter\def\csname SF@gobble@opt \endcsname{\@ifnextchar[
  \SF@gobble@twobracket
  \@gobble
}
\edef\SF@gobble@opt{\noexpand\protect
  \expandafter\noexpand\csname SF@gobble@opt \endcsname}
\def\SF@gobble@twobracket[#1]#2{}

\usepackage[T1]{fontenc}
\usepackage[latin1]{inputenc}
\usepackage{babel}
\usepackage{graphics}

\makeatletter

\usepackage[T1]{fontenc}
\usepackage[latin1]{inputenc}
\usepackage{babel}
\usepackage{graphics}

\makeatletter

\usepackage[T1]{fontenc}
\usepackage[latin1]{inputenc}
\usepackage{babel}
\usepackage{graphics}

\makeatletter

\usepackage[T1]{fontenc}
\usepackage{babel}
\usepackage{graphics}

\makeatletter

\slugcomment{}
\shorttitle{}
\shortauthors{}

\makeatother

\makeatother

\makeatother

\makeatother

\makeatother
\begin{document}

\title{Discovery of a New Nearby Star}

\author{B. J. Teegarden}

\affil{Laboratory for High Energy Astrophysics, NASA/Goddard Space Flight
Center, Greenbelt, MD 20771, US}

\email{bonnard@lheamail.gsfc.nasa.gov}

\author{S. H. Pravdo, M. Hicks, S. B. Shaklan}

\affil{Jet Propulsion Laboratory, California Institute of Technology, 4800
Oak Grove Drive, Pasadena, CA 91109, USA}

\author{K. Covey, O. Fraser, S. L. Hawley}

\affil{University of Washington, Dept. of Astronomy, Box 351589, Seattle,
WA 98195,USA}

\author{T. McGlynn}

\affil{Laboratory for High Energy Astrophysics, NASA/Goddard Space Flight
Center, Greenbelt, MD 20771, US}

\and{}

\author{I. N. Reid}

\affil{Space Telescope Science Institute, 3700 San Martin Dr., Baltimore,
MD 21218, USA}

\begin{abstract}
We report the discovery of a nearby star with a very large proper
motion of \( 5.06\pm 0.03 \) arcsec/yr. The star is called SO025300.5+165258
and referred to herein as HPMS (high proper motion star). The discovery
came as a result of a search of the SkyMorph database, a sensitive
and persistent survey that is well suited for finding stars with high
proper motions. There are currently only 7 known stars with proper
motions > 5 arcsec/yr. We have determined a preliminary value for
the parallax of \( \pi =0.43\pm 0.13 \) arcsec. If this value holds
our new star ranks behind only the Alpha Centauri system (including
Proxima Centauri) and Barnard's star in the list of our nearest stellar
neighbors. The spectrum and measured tangential velocity indicate
that HPMS is a main-sequence star with spectral type M6.5. However,
if our distance measurement is correct, the HPMS is underluminous
by \( 1.2\pm 0.7 \) mag. 
\end{abstract}


\section{Introduction}

Proper motion surveys have historically revealed a large fraction
of our nearest stellar neighbors. The usual method is to search plate
pairs typically taken a decade or more apart for relative displacements.
However, stars with large proper motions can get lost in the background.
The SkyMorph database of NEAT [Near Earth Asteroid Tracking] \citep{ref8} data is 
ideally suited to search for such objects. NEAT is a NASA-funded project intended to
perform a comprehensive survey of near-earth asteroids. Beginning
in 1995, several dedicated telescopes have been employed to repeatedly
take CCD exposures of the sky mostly in the region within \( \pm 25^{\circ } \)
of the celestial equator (excluding galactic latitudes < \( \sim 20^{\circ } \)).
The instantaneous fields-of-view have varied over the years with telescope
and CCD camera, but generally covered areas of 2-3 deg\( ^{2} \).
Our survey, thus far, has included data from May 1996 - Dec 2001 from
two 1-m telescopes located on Maui. The total sky coverage is \( \sim 10000 \)
deg\( ^{2} \), and within the region of coverage there are typically
20 exposures at any given point. The revisit time ranges from hours
to years. With such a large number of exposures over a \( \sim 6 \)
yr period, the Skymorph database is an ideal place to search for objects
with very large proper motion.~

\section{Observations}

We have searched the Skymorph database for objects with unfiltered
CCD magnitudes in the range 11.0-18.0 having proper motions > 0.25
arcsec/yr. Candidate objects were visually confirmed by comparison
with Digital Sky Survey images using the overlay feature of the SkyView
Program \citep{ref5}. A total of 3154 objects were found of which 63
had proper motions > 1.0 arcsec/yr. \( \sim 400 \) of these objects
are previously uncatalogued, and a paper on these is in preparation \citep{ref9}.
The HPMS had the largest proper motion of all the detected objects:
\( 5.06\pm 0.03 \) arcsec/yr. To further refine the HPMS proper motion
determination we have included additional observations from the USNO
A2.0 Catalogue \citep{ref6} and the 2nd Guide Star Catalogue\citep{ref7}.
The measured parameters of our fast-moving star are given in Table
1. We derive from our observations a tangential velocity of \( 52.9\pm 11.7 \)
km/s. This is consistent with the velocity dispersion of main sequence
red dwarfs of \( \sim 50 \) km/s \citep{ref1}.

Figure 1 shows the motion of HPMS across the sky with images from
the Palomar Sky Surveys and SkyMorph. The parallax was estimated by fitting the 
position of HPMS on 43 SkyMorph images (many averaged in sets of three) taken from 
Julian Dates 2450779-2452498 (November 26, 1997- August 11, 2002) to a reference 
frame defined by 12 nearby stars in the field. Since the HPMS is quite red, a 
significant correction for differential chromatic refraction (DCR) was required.
The DCR correction was both estimated from first principles and fit
to the data taken over one night with varying hour angles. We chose
a value for the DCR correction that minimized the scatter in the position
on single nights. Figure 2 shows the result of the fit to parallax
of the HPMS position with the NOVAS \citep{ref3} model for proper motion
and parallax. The result for the parallax is \( 0.43\pm 0.13 \) arcsec. The 1-sigma
error is larger than we would like because the camera pixel sizes
in the measurements, 1 and 1.4 arcsec, are not ideal for this purpose.
More precise observations with cameras designed for this purpose are
underway.

A spectrum of the HPMS taken on 11 July, 2002 using the Apache Point
2.5 m telescope is shown Figure. 3. For comparison the spectra of the M6.5 V star 
GJ1111 and the M8 V star VB10 are also shown. The match to GJ1111 is good. We identify
the HPMS as a main-sequence dwarf with spectral type M6.5 V based
upon detailed examination of the CaH2 (\( \lambda  \)7042-7046, \( \lambda  \)6814-6846),
CaH3 (\( \lambda  \)7042-7046, \( \lambda  \)6960-6990), and TiO5
(\( \lambda  \)7042-7046, \( \lambda  \)7126-7135) spectral features \citep{ref10}.
The relative strengths of the CaH and TiO bands show no evidence that 
the star is significantly metal-poor.

The HPMS was observed photometrically on the night of 7 August, 2002
at the JPL Table Mountain Observatory 0.6-meter telescope using broad
band UBVRI Bessel filters and the facility CCD imaging system. Multiple
UBVRI exposures of 15 Landolt (1992) stars ranging in color and air
mass were also measured. With this data both linear and color-corrected
extinction curves for each filter were computed. The nominal extinction
coefficients and small (\( \sim 1 \) mMag) color coefficients indicated
a good photometric night. The color-corrected extinction curves were
used to compute the measured magnitudes listed in Table 1.

The HPMS is also in the soon-to-be-released 2MASS all-sky point source
catalog [2MASS J02530084+1652532] \citep{ref4}, with magnitudes of
J = \( 8.39\pm 0.03 \), H = \( 7.88\pm 0.04 \) and K\( _{s} \)
= \( 7.59\pm 0.05 \). The photometric measurements are summarized
in Table 1. Combining these data with the trigonometric parallax given
above allows estimation of the absolute magnitude. Despite the strong
spectroscopic similarity to GJ 1111, the derived values are significantly
fainter. For example, we derive M$_J$= \( 11.58\pm 0.35 \) for the HPMS,
as compared with M$_J$ = \( 10.40\pm 0.03 \) for GJ 1111. If our trigonometric
parallax is accurate, then the HPMS is underluminous relative to a
normal M6.5 disk dwarf by \( 1.2\pm 0.7 \) magnitudes. This result
might be expected for an extreme metal-poor subdwarf, such as LHS
1826 \citep{ref12}, but, as noted above, there
is no indication that HPMS is metal-poor. If the star has the same effective 
temperature as GJ 1111, then the low luminosity implies that the HPMS has a radius 
only 60\% that of GJ 1111, or \( \sim 0.68 \) R\( _{Jupiter} \), violating 
electron degeneracy.

Given these results, we have estimated a spectrophotometric parallax
for the HPMS. \citet{ref11} have shown that spectral type is well
correlated with MJ for ultra-cool dwarfs. Using their linear relation,
we derive MJ= \( 10.59\pm 0.25 \), corresponding to a distance of
\( 3.6\pm 0.4 \) parsecs. At this distance, which is almost identical
to that of GJ 1111, HPMS is the 17th nearest system (and 27th nearest
star) to the Sun.

\section{Conclusions}

Table 2 summarizes the properties of the 10 nearest known stars including
distances and proper motions. Given our parallax value, the HPMS ranks
3rd in distance among stellar systems (or 5th among individual objects)
and 8th in proper motion. (We emphasize again that the parallax value
is preliminary and that the distance ranking could change.). In the
visual band it is the faintest star in the list, which is part of
the reason it was not discovered in previous surveys. Since the NEAT
survey has only covered the region \( \pm 25^{\circ } \) in declination
it is entirely possible that other faint nearby objects remain to
be discovered.

\acknowledgements{}

We thank D. Kirkpatrick for his reduction of the HPMS 2MASS data.
We also thank N. Zacharias, S. Levine, and C. Dahn for useful discussions.
This research was performed in part by the Jet Propulsion Laboratory,
California Institute of Technology, under contract with the National
Aeronautics and Space Administration. The results presented in this
paper are based in part on data obtained using the NASA-JPL camera
at the Oschin Telescope, Palomar Observatory, California Institute
of Technology. This publication makes use of data products from the
Two Micron All Sky Survey, which is a joint project of the University
of Massachusetts and the Infrared Processing and Analysis Center/California
Institute of Technology, funded by the National Aeronautics and Space
Administration and the National Science Foundation. SkyMorph is funded
by NASA's Applied Information Systems Program.

\clearpage

\begin{deluxetable}{ll}
\tablenum{1}
\tablecaption{HPMS Parameters}
\startdata
RA (J2000)\tablenotemark{a} &42.24369\( ^{\circ } \)\\
DEC (J2000)\tablenotemark{a} &16.89200\( ^{\circ } \)\\
Epoch\tablenotemark{a} &1989.84\\
Magnitudes (B,V, R, I, J, H, K\( _{s} \))\tablenotemark{b}&
  17.21\( \pm  \) 0.11, 15.40 \( \pm  \) 0.08, 13.26 \( \pm  \)
  .06, 10.66 \( \pm  \) 0.03,\\
  &8.39 \( \pm  \) 0.03, 7.88 \( \pm  \) 0.04, 7.59 \( \pm  \) 0.05\\ 
Spectral Type &M6.5 V\\
Proper Motion &5.06 \( \pm  \) 0.03 arcsec/yr\\
Position Angle &138.1 \( \pm  \) 0.3\( ^{\circ } \)\\
Parallax (\( \pi  \)) &0.43 \( \pm  \) 0.13 arcsec\\
Distance (1/\( \pi  \)) &2.3 (+1.0,-0.5) pc\\
Tangential Velocity &52.9 \( \pm  \) 11.7 km/s\\
\enddata

\tablenotetext{a}{From 2nd Guide Star Catalogue}
\tablenotetext{b}{B,V,R,I are from this work. J, H, K$_{s}$ are from the 2MASS
Catalogue.}

\end{deluxetable}

\begin{deluxetable}{lcccc}
\tablenum{2}
\tablecaption{Nearest Known Stars\tablenotemark{a}}
\tablehead{
\colhead{Name} & \colhead{D(pc)} & \colhead{PM\tablenotemark{b} (arcsec/yr)}
  & \colhead{M(V)\tablenotemark{c}} & \colhead{Sp. Type}}
\startdata
Proxima Centauri &1.30 &3.81 &15.49 &dM5 e\\
Alpha Centauri A &1.34 &3.69 &4.38 &G2 V\\
Alpha Centauri B &1.34 &3.69 &5.71 &K0 V\\
Barnards Star    &1.83 &10.31 &13.23 &M5 V\\
\emph{HPMS (this work)} &2.3 (+1.0,-0.5) &$5.06\pm0.03$ &$18.5\pm0.7$ &M6.5 V\\
Wolf 359         &2.39 &4.70 &16.56 &M6\\
Gl 411           &2.52 &4.81 &10.48 &M2 Ve\\
Luyten 726-8 A   &2.63 &3.37 &15.47 &dM5.5e\\
Luyten 726-8 B (UV Ceti) &2.63 &3.37 &15.60 &dM5.5e\\
Sirius A         &2.63 &1.33 &1.47 &A1 V\\
Sirius B         &2.63 &1.33 &11.34 &DA2\\
\enddata

\tablenotetext{a}{All values in this table (with the exception of the HPMS) were
either taken directly or derived from the Preliminary Version of the Third Catalogue
of Nearby Stars.{ref2}}
\tablenotetext{b}{Proper motion.}
\tablenotetext{c}{Absolute visual magnitude.}

\end{deluxetable}

\figcaption{Images from the Palomar Sky Surveys and SkyMorph (lower panels)
show the motion of HPMS south and east across the sky from 1951 until the  
current epoch.}

\figcaption{The fit to the parallax of HPMS results in a value of
0.43 \protect\protect\( \pm \protect \protect \)
0.13 arcsec. The errors per measurement that produce the good fit
shown are 0.28 arcsec. The contribution to this error from photon-counting
statistics is \protect\protect\( \sim \protect \protect \)0.1 arcsec.
The remainder is due to uncalibrated DCR or other unknown systematic 
errors.}

\figcaption{The spectrum of HPMS compared with two standard stars. Based upon
thiscomparision and detailed analysis of spectral features HPMS is a main 
sequence
dwarf with spectral type M6.5.}

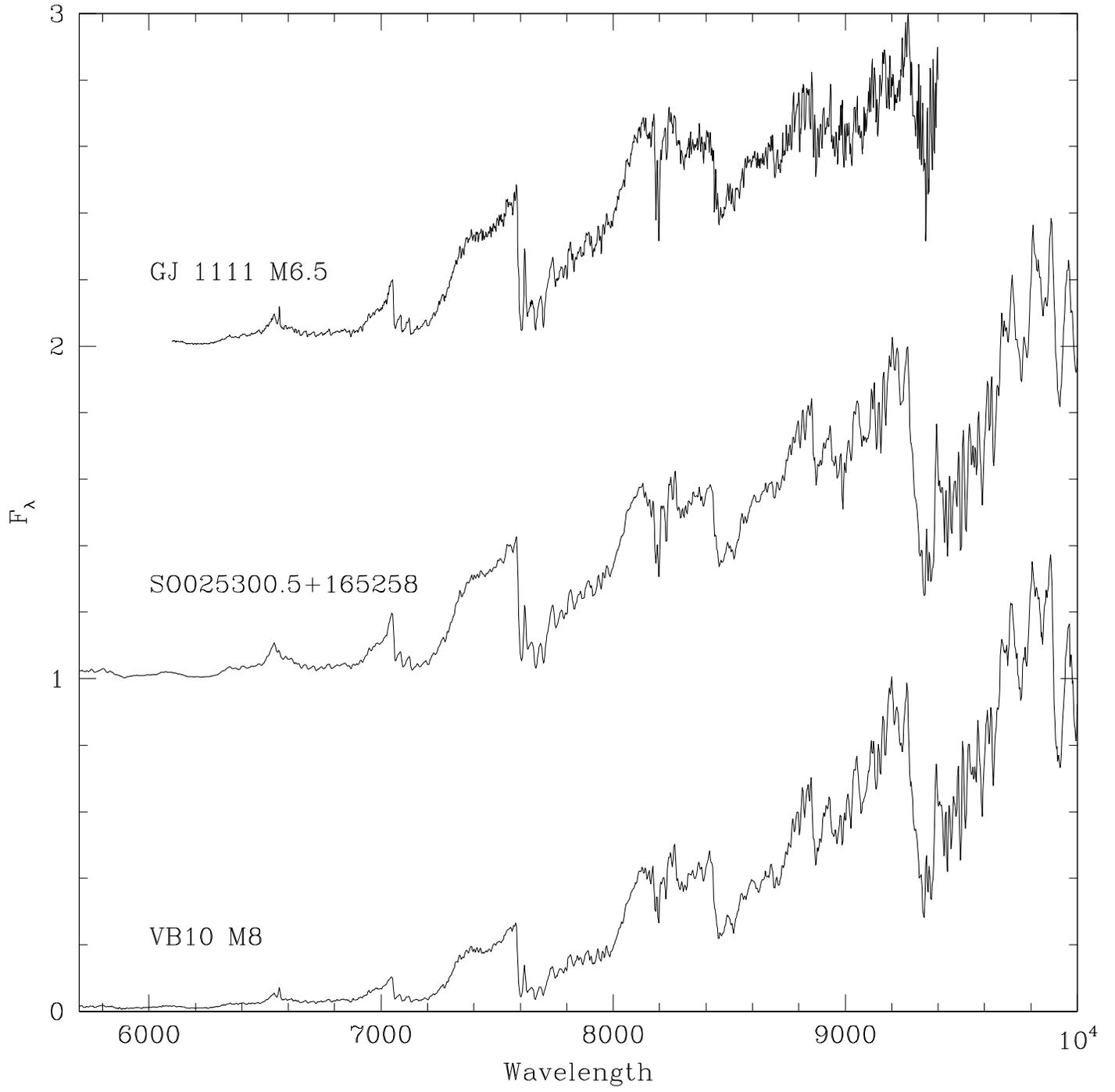
\end{document}